\documentclass[12pt,a4paper]{article}
\usepackage{latexsym}
\usepackage{graphicx}
\usepackage{subcaption}
\usepackage{amsmath,amsthm}
\usepackage{amssymb}
\usepackage{epsfig}
\usepackage{epstopdf}
\usepackage[titletoc,title]{appendix}
\usepackage{lineno}
\usepackage{float}
\usepackage{algpseudocode}
\usepackage{amsthm}
\usepackage{etoolbox}
\usepackage{longtable}
\usepackage{tikz}
\usepackage{fancyhdr}
\usepackage[T1]{fontenc}
\usepackage{rotating}
\usepackage{algorithm}
\usepackage{array}
\usepackage{makecell}
\usepackage{tabularx}
\usepackage{array}
\usepackage{graphicx}
\newtheorem{thm}{Theorem}[section]

\theoremstyle{definition}

\theoremstyle{remark}

\makeatletter \@addtoreset{equation}{section}
\makeatother
\title{A Three-Parameter Binary Subdivision Scheme for Shape-Controlled Curve Design}

\author{
	Rabia Hameed\thanks{Corresponding Author E-mail:rabia.hameed@gscwu.edu.pk} \  Jihad Younis\thanks{Corresponding Author E-mail: jihadyounis162@gmail.com} \ \ Maryam Salem Alatawi\thanks{E-mail: msoalatawi@ut.edu.sa} \ Tahreem Farman\thanks{Email: tahreemfarman330@gmail.com} \\
	\ and \  Hafiza Sana Mukhtiar\thanks{E-mail: sanamukhtar58@gmail.com}\
	\\ *$\S$$\P$ \small{Department of Mathematics, The Government Sadiq College Women University Bahawalpur,}\\ \small{Bahawalpur, Pakistan}\\
	\dag \small {Department of Mathematics, University of Aden, Aden, Yemen}\\
	\ddag \small {Department of Mathematics, Faculty of Science, University of Tabuk, Tabuk, Saudi Arabia}
}	

\date{}
\begin{document}
	\maketitle
%

\begin{abstract}
Shape-controlled curve design plays a fundamental role in computer-aided geometric design, computer graphics, and engineering applications. In this paper, we present a novel three-parameter $9$-point binary approximating subdivision scheme constructed by a weighted combination of the refinement rules of the $7$-point Lagrange and $7$-point B-spline subdivision schemes. The proposed construction employs displacement vectors between the corresponding refinement points of the parent schemes and constructs resultant vectors by combining neighbouring displacement vectors through three independent design-control parameters. These resultant vectors are subsequently used to derive the refinement rules, thereby yielding a unified family of binary subdivision schemes with adjustable geometric characteristics while preserving the approximating nature of the refinement process. Two representative sub-schemes are derived as special cases by imposing suitable constraints on the design-control parameters. Theoretical investigations establish the support, continuity, endpoint rules for open polygons, and Gibbs oscillation behavior of the proposed family, while graphical examples demonstrate the influence of the design-control parameters on the generated limit curves. Owing to its flexible geometric construction, intuitive parameterization, and favorable mathematical properties, the proposed subdivision framework provides a flexible and effective framework for design-controlled curve design and a valuable tool for computer-aided geometric design and related engineering applications.
\end{abstract}

	
	
	
	
	
\section{Introduction}

Subdivision schemes have become one of the most fundamental tools in computer-aided geometric design (CAGD) owing to their ability to generate smooth curves and surfaces from an initial control polygon through repeated refinement. Their mathematical simplicity, computational efficiency, and local refinement characteristics have led to widespread applications in geometric modeling, computer graphics, reverse engineering, image processing, animation, and isogeometric analysis. Depending on the relationship between the control polygon and the generated limit curve, subdivision schemes are broadly classified into interpolatory and approximating schemes. Among these two classes, approximating subdivision schemes are particularly attractive for practical geometric modeling because they generally produce smoother limit curves, possess superior approximation properties, and provide greater flexibility for geometric design.

The theoretical foundations of approximating subdivision schemes were established through several pioneering contributions that continue to influence modern subdivision theory. The refinement process introduced by de Rham in the 1950s and the corner-cutting algorithm proposed by Chaikin in 1974 demonstrated that repeated local refinement could efficiently generate smooth curves from discrete control points, thereby laying the foundation for subsequent developments in subdivision-based geometric modeling. Building upon these early ideas, Dyn et al.~\cite{Dyn1} showed that suitable perturbations of subdivision rules can improve the smoothness of limit curves while preserving convergence, establishing an important relationship between refinement masks and geometric continuity. Dyn~\cite{Dyn2221} later presented a comprehensive treatment of subdivision algorithms and multiresolution techniques, highlighting both their mathematical properties and practical significance in geometric modeling. The analytical framework of subdivision theory was further strengthened by Levin~\cite{Levin}, who introduced the Laurent polynomial formulation for analyzing non-uniform binary subdivision schemes, a methodology that has become one of the standard tools for investigating convergence and smoothness. In addition, Ivrissimtzis et al.~\cite{Ivrissimtzis} investigated the support of recursive subdivision schemes and clarified the locality of refinement, providing valuable insight into the geometric influence of subdivision masks on the resulting limit curves.

These foundational developments have established the theoretical basis upon which modern approximating subdivision schemes are constructed. Consequently, subsequent research has focused not only on improving smoothness and approximation properties but also on developing subdivision schemes with enhanced geometric flexibility, compact support, higher continuity, and systematic construction methodologies suitable for increasingly sophisticated geometric modeling applications.

Building upon these theoretical foundations, considerable research has been devoted to improving the smoothness, approximation capability, and computational efficiency of binary approximating subdivision schemes. Among the notable contributions, Siddiqi and Ahmad~\cite{Siddiqi1001} proposed a five-point binary approximating subdivision scheme capable of generating $C^{4}$ continuous limit curves and subsequently extended their work to a stationary scheme that achieves $C^{6}$ continuity while maintaining compact support~\cite{Siddiqi2021}. Daniel and Shunmugaraj~\cite{Daniel} introduced a parameterized three-point approximating subdivision scheme whose special cases include several well-known B-spline curves. A broader perspective was presented by Mustafa et al.~\cite{Mustafa2002}, who established a unified family of $(2n-1)$-point binary approximating subdivision schemes, thereby providing a systematic framework that encompasses numerous existing constructions. Collectively, these studies have substantially advanced the development of binary approximating subdivision schemes with higher continuity, compact support, and improved approximation behavior.

Alongside the pursuit of higher smoothness, increasing attention has been directed toward enhancing the geometric flexibility of subdivision schemes through the incorporation of design-control parameters. In this context, Mustafa et al.~\cite{Mustafa5} introduced a unified three-point approximating subdivision scheme governed by three independent design-control parameters, while Mustafa and Randhawa~\cite{Mustafa1} carried out its complete theoretical investigation by analyzing convergence, continuity, and support. Parameterized refinement strategies have subsequently been explored from several complementary perspectives. Ghaffar et al.~\cite{Ghaffar} proposed a four-point $\alpha$-array approximating subdivision scheme in which a shape parameter provides direct control over the resulting limit curves, and later generalized this concept to subdivision schemes of varying arity with reduced support and improved approximation properties~\cite{Ghaffar3}. Fang et al.~\cite{Fang1} developed a family of non-uniform subdivision schemes with variable parameters, whereas Tan et al.~\cite{Tan} presented parameterized five-point and shape-preserving four-point subdivision schemes that further enhanced local geometric control. More recently, Hameed et al.~\cite{Hameed2023} introduced a design-control approximating refinement scheme that demonstrates how suitable parameterization can generate a diverse range of smooth curve profiles while preserving the convergence and smoothness characteristics of the underlying subdivision process. These contributions clearly indicate that design-control parameters have become an effective mechanism for improving the adaptability of approximating subdivision schemes without compromising their desirable mathematical properties.

The growing complexity of geometric modeling problems has also motivated the development of systematic construction methodologies capable of generating entire families of subdivision schemes from unified mathematical frameworks. Rather than deriving individual refinement rules independently, these approaches seek to establish general principles that simplify the construction of new subdivision schemes while preserving desirable theoretical properties. Zheng et al.~\cite{Zheng} proposed a general formulation for integer-point binary approximating subdivision schemes, whereas Shi et al.~\cite{Shi} constructed six-point subdivision schemes with cubic precision and established their relationship with quintic B-spline refinement. Alternative unified construction strategies were developed by Asghar and Mustafa through Laurent polynomial formulations for deriving families of $a$-ary subdivision schemes~\cite{Asghar1201} and a binomial distribution-based framework for binary approximating subdivision schemes~\cite{Asghar121}. More recently, generalized construction techniques have been extended beyond binary refinement. Nosheen et al.~\cite{Nosheen2024} established a unified relationship between even-point binary subdivision schemes and their even- and odd-point quaternary counterparts. Building upon this concept, Hameed et al.~\cite{Hameed2025SciRep} proposed a generalized framework for converting $(2n+1)$-point binary subdivision schemes into $(3n+1)$-point quaternary schemes, while Hameed and Mustafa~\cite{Hameed2025RIM} introduced a unified methodology for constructing relaxed quaternary subdivision schemes from relaxed binary schemes. These studies collectively demonstrate that generalized construction frameworks provide an effective means of systematically generating new subdivision schemes while preserving the essential characteristics of their parent formulations.

In parallel with these developments, considerable effort has been devoted to enhancing the geometric quality of subdivision-generated curves. One important direction concerns the preservation of desirable geometric properties, such as monotonicity, convexity, curvature, and torsion, which are essential for producing visually pleasing and physically meaningful geometric models. Bibi et al.~\cite{Bibi} investigated the influence of tension parameters on monotonicity and convexity preservation in binary approximating subdivision schemes, whereas Hussain et al.~\cite{Hussain} generalized five-point approximating subdivision schemes of different arities and analyzed their convergence, continuity, support, and approximation behavior. Ghaffar et al.~\cite{Ghaffar313} developed a family of $2m$-point binary non-stationary subdivision schemes possessing attractive geometric characteristics, including curvature, torsion, and convexity preservation. Similarly, Hameed and Mustafa~\cite{Hameed} proposed a family of $a$-point $b$-ary subdivision schemes with bell-shaped refinement masks, providing additional flexibility for shape-controlled geometric design.

Another active line of research addresses the reduction of oscillatory artifacts while maintaining the smoothness of subdivision-generated curves. Amat et al.~\cite{Amat1} investigated the Gibbs phenomenon in stationary subdivision schemes based on the classical analysis of Gibbs oscillations by Gottlieb and Shu~\cite{Gottlieb7}. Their work was subsequently extended by Zhou et al.~\cite{Zhou111} to $p$-ary subdivision schemes, while Amat et al.~\cite{Amat2} developed nonlinear non-oscillatory subdivision schemes with arbitrary regularity and established their convergence, stability, and approximation properties. A related piecewise nonlinear refinement strategy was proposed by Hameed et al.~\cite{Hameed2024}, who demonstrated improved curve representations through nonlinear binary subdivision. More recently, Hameed et al.~\cite{Hameed2024Results} introduced a seven-point subdivision scheme equipped with design-control parameters that simultaneously achieves high smoothness and reduced Gibbs oscillations. Complementing these studies, Dyn et al.~\cite{Dyn33} showed that non-uniform interpolatory subdivision schemes can attain improved smoothness without increasing the support size, highlighting the continued evolution of subdivision methodologies toward increasingly efficient and geometrically robust refinement algorithms.

Despite the significant progress achieved in approximating subdivision schemes, several challenges remain. Existing studies have successfully improved continuity, enhanced approximation properties, incorporated design-control parameters, and developed generalized construction methodologies for generating new subdivision schemes. Nevertheless, most high-order binary approximating schemes have been derived through direct algebraic manipulation of refinement masks, with comparatively less emphasis on exploiting the geometric relationships between existing subdivision schemes as a systematic design strategy. Moreover, although parameterized subdivision schemes have substantially improved geometric flexibility, many available formulations employ only one or two free parameters, thereby restricting the range of attainable curve shapes and limiting local control during the design process. These observations suggest that there is still considerable scope for developing subdivision schemes that combine a simple geometric construction procedure with enhanced design flexibility while preserving desirable theoretical properties.

The demand for such subdivision schemes is further reinforced by modern applications in computer-aided geometric design, reverse engineering, digital manufacturing, computer graphics, and isogeometric analysis, where smooth and shape-controllable curves are essential for accurately representing complex geometries. In these applications, refinement algorithms are expected not only to possess strong mathematical characteristics, such as convergence, compact support, and high-order continuity, but also to provide sufficient flexibility to satisfy diverse design requirements without increasing computational complexity. Consequently, constructing subdivision schemes that simultaneously achieve high smoothness, effective design control, and computational efficiency remains an important and active research objective.

Motivated by these considerations, this paper presents a novel three-parameter $9$-point binary approximating subdivision scheme constructed through a geometric combination of the classical $7$-point Lagrange and $7$-point B-spline subdivision schemes. Unlike conventional approaches that derive refinement rules solely through algebraic formulations, the proposed methodology first evaluates the displacement vectors between the corresponding refinement points of the two parent schemes and subsequently combines neighboring displacement vectors to obtain resultant vectors that determine translated refinement points. This geometric construction naturally produces a unified family of binary subdivision schemes governed by three independent design-control parameters, enabling enhanced local curve modification while preserving the approximating nature, convergence, and smoothness of the refinement process.

The proposed framework also encompasses representative one- and two-parameter subdivision schemes as special cases through suitable restrictions on the design-control parameters. A comprehensive theoretical analysis is presented to investigate the support, continuity, and endpoint refinement rules for open control polygons, together with the Gibbs oscillation behavior of the proposed subdivision family. Finally, several numerical examples are provided to demonstrate the influence of the design-control parameters on the generated limit curves and to illustrate the effectiveness of the proposed scheme for shape-controlled curve design in computer-aided geometric design and related engineering applications.

The remainder of this paper is organized as follows. Section~2 reviews the preliminary concepts and mathematical tools used throughout the paper. Section~3 presents the geometric construction of the proposed three-parameter 9-point binary subdivision scheme together with its representative sub-schemes. Section~4 investigates the support of the proposed subdivision family and presents the endpoint refinement rules for open and closed control polygons. Section~5 establishes the continuity properties of the proposed schemes. Section~6 examines the Gibbs oscillation behavior of the proposed schemes, and Section~7 concludes the paper by summarizing the main findings and discussing directions for future research.

\section{Preliminaries}

A binary subdivision scheme generates a sequence of refined control polygons by repeatedly applying a refinement rule to an initial control polygon. Let $P^{k}=\{p_i^{k}\}_{i\in\mathbb{Z}}$ denote the control polygon at the $k$th refinement level. The refinement process is governed by a finite mask $\{a_i\}_{i\in\mathbb{Z}}$ and is expressed as
\begin{align}
	p_{i}^{k+1}=\sum_{j\in\mathbb{Z}}a_{i-2j}p_j^k,
	\tag{2.1}
\end{align}
where $a_i$ are the mask coefficients. The corresponding Laurent polynomial associated with the subdivision mask is
\begin{align}
	a(z)=\sum_{i\in\mathbb{Z}}a_i z^i,
	\tag{2.2}
\end{align}
which provides a convenient representation for analyzing the convergence and smoothness properties of subdivision schemes.

The proposed construction is based on the refinement rules of the classical $7$-point Lagrange and $7$-point B-spline subdivision schemes. For a given set of distinct data points
$\{(x_0,y_0),(x_1,y_1),\ldots,(x_n,y_n)\}$,
the Lagrange interpolation polynomial is defined as
\begin{equation}\label{eq:lagrange}
	p(x)=\sum_{k=0}^{n}L_{n,k}(x)y_k,
\end{equation}
where
\begin{equation}
	L_{n,k}(x)=
	\prod_{\substack{i=0\\ i\neq k}}^{n}
	\frac{x-x_i}{x_k-x_i},
\end{equation}
and
\[
L_{n,k}(x_i)=\delta_{ik},
\]
where $\delta_{ik}$ denotes the Kronecker delta. The interpolation property of the Lagrange basis functions makes them well suited for constructing interpolatory refinement rules.

In contrast, a B-spline curve of degree $k$ is represented by
\begin{equation}\label{eq:bspline}
	B(t)=\sum_{i=0}^{n}N_{i,k}(t)p_i,
\end{equation}
where $N_{i,k}(t)$ denotes the $i$th B-spline basis function of degree $k$. Owing to their compact support and high smoothness, these basis functions provide excellent local control over the generated curves and form the foundation of many approximating subdivision schemes.

For a binary subdivision scheme with refinement mask $\{a_i\}_{i\in\mathbb{Z}}$, a necessary condition for convergence is
\begin{equation}\label{eq:conv}
	\sum_{j\in\mathbb{Z}}a_{2j}=1,
	\qquad
	\sum_{j\in\mathbb{Z}}a_{2j+1}=1.
\end{equation}

The following theorem due to Shi \emph{et al.}~\cite{Shi} provides a sufficient criterion for establishing the continuity of binary subdivision schemes.

\begin{thm}
	Let $S$ be a subdivision scheme with mask
	$a^{(0)}=\{a_i^{(0)}\}_{i\in\mathbb{Z}}$.
	Suppose that its $j$th-order divided difference schemes
	$S_j$ $(j=1,\ldots,n+1)$
	exist with masks
	$a^{(j)}=\{a_i^{(j)}\}_{i\in\mathbb{Z}}$
	satisfying
	
	\[
	\sum_{i\in\mathbb{Z}}a_{2i}^{(j)}
	=
	\sum_{i\in\mathbb{Z}}a_{2i+1}^{(j)}
	=
	1,
	\qquad
	j=0,1,\ldots,n.
	\]
	
	If there exists an integer $L\ge1$ such that
	
	\[
	\left\|
	\left(\frac12S_{n+1}\right)^L
	\right\|_\infty<1,
	\]
	
	then the subdivision scheme $S$ is $C^n$-continuous, where
	
	\[
	\left\|
	\left(\frac12S_{n+1}\right)^L
	\right\|_\infty
	=
	\max
	\left\{
	\sum_{j\in\mathbb{Z}}
	\left|
	b_{i-2^Lj}^{[L]}
	\right|
	:
	0\le i<2^L
	\right\},
	\]
	
	with
	
	\[
	b^{[L]}(z)
	=
	b(z)b(z^2)\cdots b(z^{2^{L-1}}),
	\qquad
	b(z)=\frac12a^{(n+1)}(z).
	\]
	
	In particular, when $L=1$,
	
	\[
	\left\|
	\frac12S_{n+1}
	\right\|_\infty
	=
	\frac12
	\max
	\left\{
	\sum_{i\in\mathbb{Z}}
	|a_{2i}^{(n+1)}|,
	\,
	\sum_{i\in\mathbb{Z}}
	|a_{2i+1}^{(n+1)}|
	\right\}.
	\]
	
	Moreover, if
	
	\[
	a(z)=\frac{(1+z)^{n+1}}{2^n}b(z),
	\]
	
	where $S_b$ is contractive, then $S_a$ is convergent and
	$C^n$-continuous.
\end{thm}

The Gibbs phenomenon plays an important role in assessing the visual quality of subdivision-generated curves near discontinuities. The following theorem established by Amat \emph{et al.}~\cite{Amat1} provides a useful criterion for analyzing the oscillatory behavior of stationary subdivision schemes.

\begin{thm}\label{a00}
	Let $0\le\xi\le h$, and let
	
	\[
	f(x)=
	\begin{cases}
		f_{-}(x), & x\le\xi,\\
		f_{+}(x), & x>\xi,
	\end{cases}
	\]
	
	where
	$f_{-}\in C^n((-\infty,\xi])$,
	$f_{+}\in C^n((\xi,\infty))$,
	$n\ge2$,
	and
	$f_{-}(\xi)>f_{+}(\xi)$.
	Define
	
	\[
	\zeta_l^{[k]}(i)=
	\begin{cases}
		\displaystyle
		\sum_{\tau\le i}a_{2^k\tau+l}^{[k]}, & i<0,\\[2mm]
		0, & i=0,\\[2mm]
		\displaystyle
		\sum_{\tau\ge i}a_{2^k\tau+l}^{[k]}, & i>0,
	\end{cases}
	\]
	
	where $a$ denotes the subdivision mask and
	$0\le l<2^k$.
	If
	$\zeta_l^{[k]}(i)\ge0$
	for all $i$ and $k$, and $h$ is sufficiently small, then the subdivision limit satisfies the approximation and Gibbs bounds established in~\cite{Amat1}.
\end{thm}

\section{Construction of a 9-point binary subdivision scheme via a weighted combination of two 7-point subdivision schemes}

This section presents the construction of a novel binary $9$-point approximating subdivision scheme by combining the refinement rules of two established $7$-point binary subdivision schemes, namely the Lagrange approximating scheme and the B-spline approximating scheme. The proposed formulation introduces three design-control parameters that provide additional degrees of freedom for local shape modification while preserving the approximating nature of the refinement process. By appropriately adjusting these parameters, a broad family of subdivision schemes with enhanced geometric flexibility can be generated.

The $7$-point Lagrange approximating subdivision scheme is defined by
\begin{equation}\label{a1}
	\left\{
	\begin{aligned}
		Q_{2i,7}^{k+1} &= \sum_{j=-3}^{3} a_j p_{i+j,7}^{k},\\
		Q_{2i+1,7}^{k+1} &= \sum_{j=-3}^{3} a_{-j} p_{i+j,7}^{k},
	\end{aligned}
	\right.
\end{equation}
where
\begin{align*}
	a_{-3}&=\frac{273}{65536}, &
	a_{-2}&=-\frac{1287}{32768}, &
	a_{-1}&=\frac{15015}{65536},\\
	a_{0}&=\frac{15015}{16384}, &
	a_{1}&=-\frac{9009}{65536}, &
	a_{2}&=\frac{1001}{32768},\qquad
	a_{3}=-\frac{231}{65536}.
\end{align*}

Similarly, the $7$-point B-spline approximating subdivision scheme is given by

\begin{equation}\label{a2}
	\left\{
	\begin{aligned}
		R_{2i,7}^{k+1} &= \sum_{j=-3}^{3} b_j p_{i+j,7}^{k},\\
		R_{2i+1,7}^{k+1} &= \sum_{j=-3}^{3} b_{-j} p_{i+j,7}^{k},
	\end{aligned}
	\right.
\end{equation}
where
\begin{align*}
	b_{-3}&=\frac{13}{4096}, &
	b_{-2}&=\frac{143}{2048}, &
	b_{-1}&=\frac{1287}{4096},\\
	b_{0}&=\frac{429}{1024}, &
	b_{1}&=\frac{715}{4096}, &
	b_{2}&=\frac{39}{2048},\qquad
	b_{3}=\frac{1}{4096}.
\end{align*}

\begin{figure}[htb]
	\centering
	\begin{tabular}{cc}
		\epsfig{file=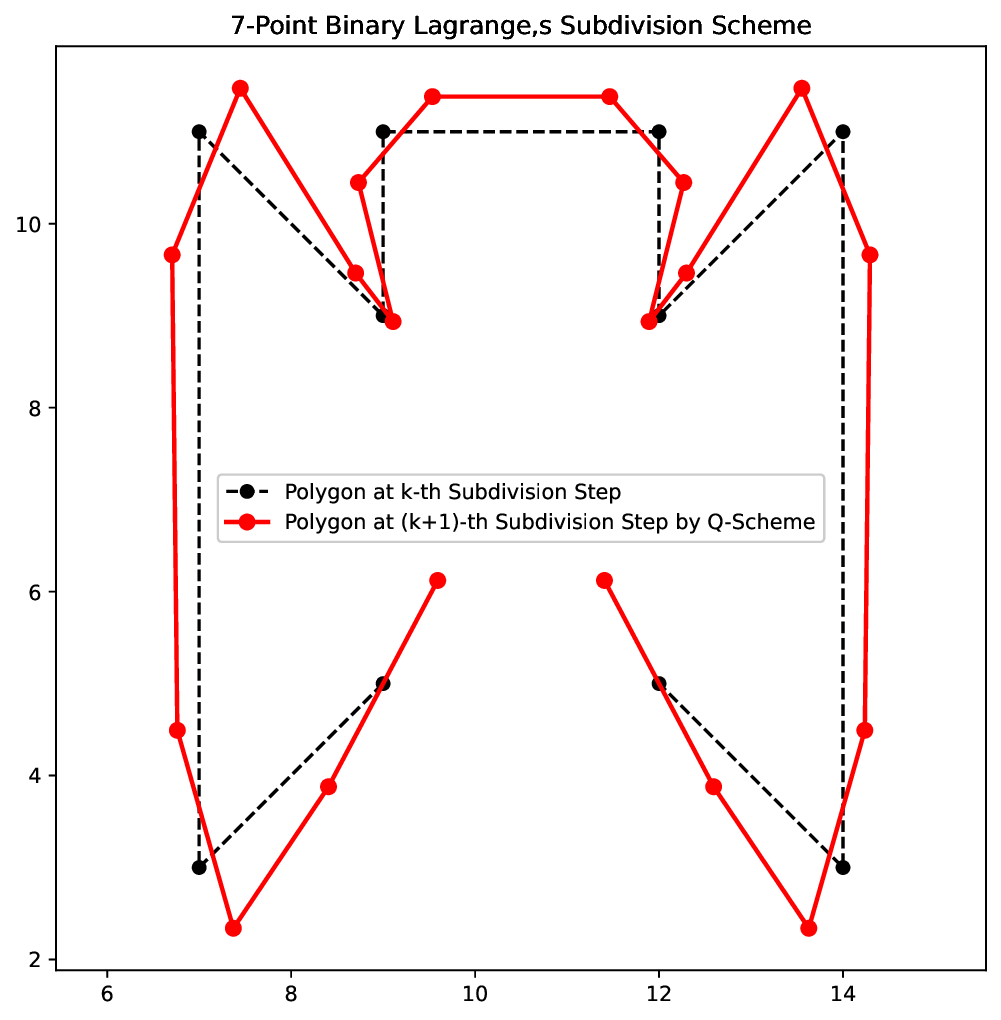,width=2.8in} &
		\epsfig{file=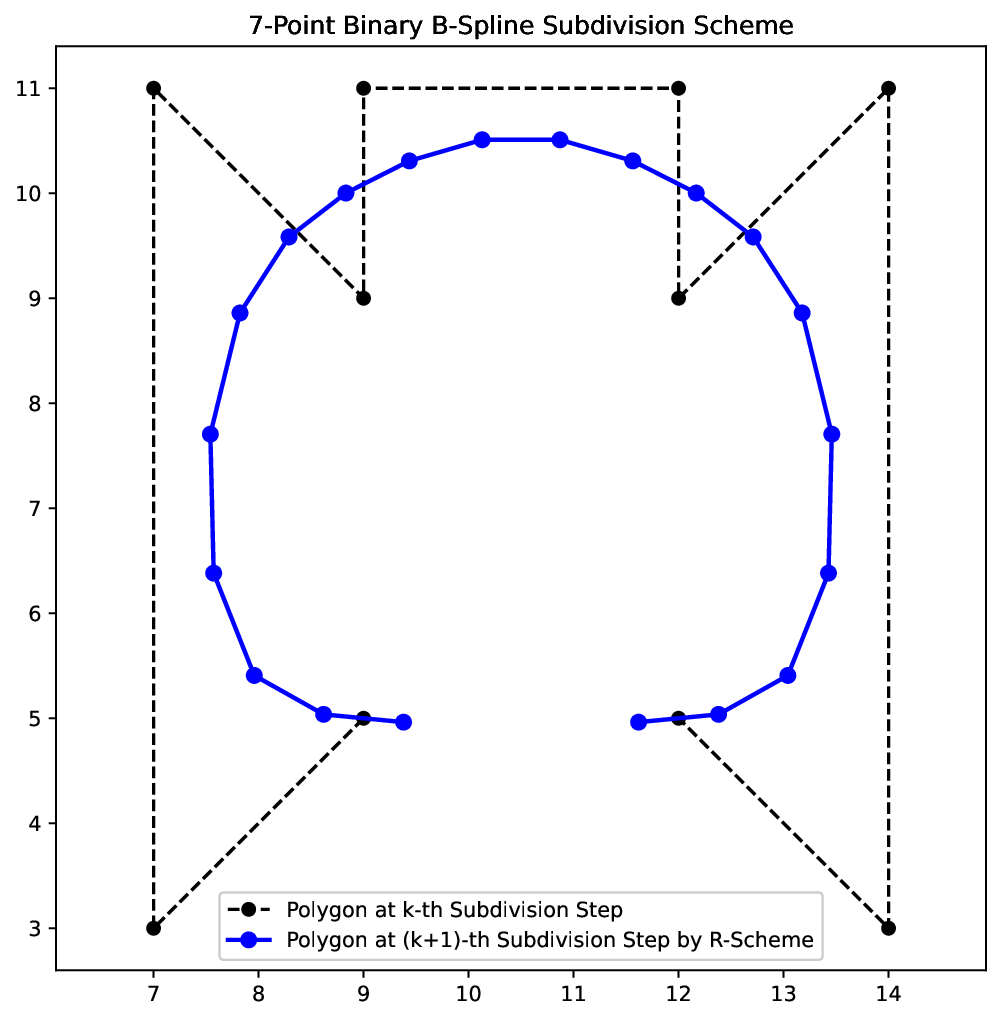,width=2.8in}\\
		(a) Q-scheme & (b) R-scheme
	\end{tabular}
\caption{Comparison of the first refined control polygons generated by the classical $7$-point Lagrange ($Q$) and $7$-point B-spline ($R$) subdivision schemes from the same initial control polygon.}
	\label{Q-R-Schemes}
\end{figure}

Figure~\ref{Q-R-Schemes} illustrates the first refined control polygons generated by the two parent subdivision schemes from the same initial control polygon.

As illustrated in Figure~\ref{Q-R-Schemes}, the two parent schemes exhibit distinct geometric characteristics. The B-spline scheme is based on a convex refinement mask, producing smooth approximating curves that remain within the convex hull of the control polygon. In contrast, the Lagrange scheme employs an affine refinement mask containing negative coefficients, which improves polynomial reproduction and preserves local geometric features more effectively, although the resulting curve may extend beyond the convex hull. The proposed $9$-point subdivision scheme combines these complementary properties to achieve a better balance between geometric fidelity, local shape control, and curve smoothness.
				
The displacement vectors between the corresponding refined points generated by the Lagrange and B-spline subdivision schemes are defined as
\[
\vec{V}_{2i,7}^{k+1}=Q_{2i,7}^{k+1}-R_{2i,7}^{k+1},
\qquad
\vec{V}_{2i+1,7}^{k+1}=Q_{2i+1,7}^{k+1}-R_{2i+1,7}^{k+1},
\]
where $Q_{2i+r,7}^{k+1}$ and $R_{2i+r,7}^{k+1}$ ($r=0,1$) are given by (\ref{a1}) and (\ref{a2}), respectively. Substituting (\ref{a1}) and (\ref{a2}) yields
\begin{equation}\label{a3}
	\left\{
	\begin{aligned}
		\vec{V}_{2i,7}^{k+1}
		&=\sum_{j=-3}^{3} c_j p_{i+j,7}^{k},\\
		\vec{V}_{2i+1,7}^{k+1}
		&=\sum_{j=-3}^{3} c_{-j} p_{i+j,7}^{k},
	\end{aligned}
	\right.
\end{equation}
where
\[
c_{-3}=-\frac{247}{65536},\;
c_{-2}=\frac{377}{32768},\;
c_{-1}=-\frac{20449}{65536},\;
c_{0}=\frac{8151}{16384},\;
c_{1}=-\frac{5577}{65536},\;
c_{2}=-\frac{3575}{32768},\;
c_{3}=\frac{65}{65536}.
\]

\section{Conclusion}

In this paper, we presented a novel three-parameter $9$-point binary approximating subdivision scheme constructed through a weighted combination of the refinement rules of the classical $7$-point Lagrange and $7$-point B-spline subdivision schemes. The proposed geometric construction utilizes displacement vectors between the corresponding refinement points of the parent schemes and combines neighbouring displacement vectors through three independent design-control parameters. The resulting resultant vectors are subsequently employed to translate the refinement points of the Lagrange scheme, thereby generating a unified family of binary subdivision schemes with enhanced geometric flexibility.

Two representative sub-schemes were derived by imposing suitable constraints on the design-control parameters, demonstrating that the proposed formulation provides a unified framework encompassing several important parameterized subdivision schemes. Theoretical investigations established the support, endpoint refinement rules for open control polygons, continuity, and Gibbs oscillation behavior of the proposed subdivision family. Furthermore, graphical examples illustrated the influence of the design-control parameters on the generated basis functions and limit curves, while numerical experiments verified the compact local support property of the proposed refinement scheme.

The proposed framework combines geometric flexibility with favorable mathematical properties while preserving the approximating nature of the refinement process. The proposed geometric construction provides a systematic approach for generating parameterized binary subdivision schemes with enhanced curve design flexibility. Owing to its compact support, high smoothness, and effective local shape control, the proposed subdivision family is well suited for computer-aided geometric design, geometric modeling, and related engineering applications. Future work will focus on extending the proposed geometric construction strategy to subdivision schemes of higher arity and to subdivision schemes for surface generation.

\section*{Author Contributions}

\textbf{Rabia Hameed} conceived and designed the study, developed the proposed methodology, carried out the theoretical and numerical analyses, interpreted the results, and wrote the original manuscript. \textbf{Jihad Younis} supervised the research and critically reviewed the manuscript. \textbf{Maryam Salem Alatawi} contributed to the interpretation of the results and manuscript revision. \textbf{Tahreem Farman} and \textbf{Hafiza Sana Mukhtiar} equally contributed to the numerical implementation, verification of the theoretical results, figure preparation, literature review, and manuscript editing. All authors read and approved the final manuscript.

\section*{Disclosure statement}
The authors declare that there are no conflicts of interest regarding the publication of this paper.


\section*{ORCID}
Rabia Hameed: https://orcid.org/0000-0003-4246-8253 \\
Jihad Younis: http://orcid.org/0000-0001-7116-3251\\
Maryam Salem Alatawi:
http://orcid.org/0000-0002-5567-7444
\section*{Data Availability}
No datasets were generated or analyzed during the current study.

\section*{Funding declaration}
The authors declare that no funds, grants, or other financial support were received for this research.

\end{document}